\documentstyle[aps]{revtex}
\begin{document}
\title{Magnetic-Field-Controlled Twin Boundaries Motion and Giant
Magneto-Mechanical Effects in Ni-Mn-Ga Shape Memory Alloy}
\author{A. A. Likhachev}
\address{Institute of Metal Physics, Dept. of Phase Transitions, Vernadsky St., 36,\\
252142 Kiev, Ukraine.}
\author{K. Ullakko}
\address{Helsinki University of Technology, Dept. of Engineering Physics and\\
Mathematics, Rakentajankio 2C, 02150 Espoo, Finland}
\maketitle

\begin{abstract}
This report represents some new experimental results and the quantitative
model describing large magneto-strain effect and main mechanical and
magnetic properties observed in several ferromagnetic shape-memory alloys.
The model application to giant magneto-strain effect recently found in some
non-stoichiometric Ni-Mn-Ga alloys is discussed.
\end{abstract}



\section{Introduction}

Some ferromagnetic shape memory alloys were recently suggested as a general
way for the development of a new class of the magnetic-field-controlled
actuator materials that will allow control of large strain effect by
application of a magnetic field. Numerous candidate shape memory materials
were explored including Ni$_2$MnGa Co$_2$MnGa, FePt CoNi, and FeNiCoTi
during past few years. Large magnetically driven strain effects are expected
have to occur in these systems. References [8-24] contain a detailed
information on this subject. Up to date, the largest magneto-strain effects
were achieved in Ni-Mn-Ga ferromagnetic shape memory alloys [1-7]. The best
results on the field-induced strain reported short time ago [2] never
exceeded 0.2-0.3\% strain value in a magnetic field of order 8 kOe in
martensitic state for nearly stoichiometric single-crystal samples of Ni$_2$%
MnGa alloy. Recently, several research groups [6,7] have reported on the
observation of super-large more than 5\% magneto-strain effect in some
non-stoichiometric Ni-Mn-Ga alloys close to that 5.78\% value expected from
the tetragonality aspect ratio of the martensite crystal lattice. New
Ni-Mn-Ga alloys showing giant magneto-strain effect display simultaneously
few interesting physical effects and new behavior for some magnetic and
mechanical properties which is very different from that earlier observed in
Ni$_2$MnGa with lower (%
\mbox{$<$}
0.3\AA ) magneto-strain effect.

One the new non-stoichiometric Ni$_{48}$Mn$_{30}$Ga$_{22}$ alloy has
recently been studied in details. This alloy undergoes the martensitic
transformation at about 35C$^0$ from the L$_{21}$ ordered Heusler type cubic
(a = 0.582 nm) crystal structure into a tetragonally distorted structure
with crystalline lattice parameters: a=b=0.594 nm and c=0.562 nm. The
martensitic phase is ferromagnetic one and internally twinned. It contains a
mixture of twin variants twinned usually on $\left\{ {110}\right\} $ planes.
The orientation of their c-axes is nearly parallel to [100] directions of
the parent cubic phase. Tetragonal symmetry direction is simultaneously a
single easy magnetization axis for each martensitic variant. Very low 2-3
MPa compressive twinning stress along [100] directions needed to transform
the multi-variant samples into a single-variant state has been found for
Ni-Mn-Ga alloys displaying giant field-induced deformation effect. It is
much less than it was observed in some earlier studied alloys (15-20 MPa
[8,9]). As a result, new alloys which usually have very simple two-variant
twinning microstructure and very low twinning stresses can be easily
transformed between two single twin variants by application both the
mechanical stress and the magnetic field through the mechanism of twin
boundaries motion.

For instance, Fig.1 shows an example of giant 5\% magneto-strain effect
associated with the direct and reverse magnetic-field-induced transformation
between two single twin variants in Ni$_{48}$Mn$_{30}$Ga$_{22}$ sample. The
magneto-strain value achieved in this new alloy is insufficiently less
compared to the natural crystallographic limit 1-c/a = 5.78\% expected for
twinning in this material.

The next Fig.2 represents in situ observation of the transformation process
between two twin variants of martensitic phase in the same Ni-Mn-Ga sample
through the mechanism of the magnetic-field-controlled twin boundary motion.
These experiments directly confirm the magnetic-field-induced twinning as
the main physical mechanism responsible for a super-large magneto-strain
effect found in this material.

Simultaneously, some important steps were made during the past few years
concerning the physical nature of the magneto-mechanical phenomena observed
in ferromagnetic shape memory alloys. James and Wuttig [4] developed the
models based on a constrained magnetization orientation approach. O'Handley
[5] first included the rotation magnetization effects together with usual
Zeeman energy term into consideration. Likhachev and Ullakko [3] have
developed a general thermodynamic approach based on the Maxwell's
relationships and derived the main magneto-mechanical equation representing
both the magnetic and the mechanical driving force balance in ferromagnetic
shape memory materials. The last model has been also successfully applied to
explain quantitatively the results of experimental study of large
magneto-strain effects in Ni$_2$MnGa earlier obtained in [2]. In particular,
according to our model estimation made in [3] a giant magneto-strain values
can be achieved only in materials with very low (%
\mbox{$<$}
2 MPa) twinning stress values. Mechanical testing results obtained for giant
magneto-strain Ni-Mn-Ga samples have completely confirmed these estimations.

The present report represents some experimental results and several new
effects recently found in some non-stoichiometric Ni-Mn-Ga alloys that are
directly connected with their giant magneto-strain response. We also
represent some new model background based on the magnetization free energy
consideration that allows better understand its important role. Naturally,
this approach gives the same final equations derived in our preceding
publications. Finally, the model equations are applied to explain some new
experimental results.

\section{Magnetization free energy and magnetic driving forces}

Perhaps, starting from a simplest physical idea first clearly formulated in
[1], it is generally believed that a large macroscopic mechanical strain
induced by the magnetic field in some ferromagnetic shape memory alloys is
usually realized trough the twin boundaries motion and redistribution of
different twin variant fractions in a magnetic field. In this case, the main
thermodynamic driving forces should have a magnetic nature and be connected
with high magnetization anisotropy and significant differences in
magnetization free energies for different twin variants of martensite [1-7].
It should be noted that some early models have considered each twin band as
a single magnetic domain with a constant magnetization value and constrained
magnetization direction. This approach predicted some unlimited linear
growth of the magnetic driving force in a magnetic field applied.
Respectively, it was expected that one can easily obtain significant
magneto-strain effects in ferromagnetic shape memory alloys trough the twin
boundary motion simply increasing magnetic field value as high as needed. It
has been understood later [5, 3] that it is not so, because each twin band
is not a single magnetic domain but should contain the multiple internal
magnetic domain microstructure and typical expected for the uniaxial
ferromagnet 180$^0$ domain wall configurations. As a result, the local
average magnetization inside of twins is not a constant and can be changed
in a magnetic field due to both the magnetic domain wall motion and the
rotation of magnetization. The most important conclusion following from this
point of view is that the magnetic driving force can not be increased
infinitely in a magnetic field but always limited its maximal value
independent on the magnetic field and proportional the magnetic anisotropy
constant. This value is low enough, so only the very soft (in sense of
twinning stress) materials can be good candidates to show really high
magnetic field induced strain effects. Therefore, the investigation of
magnetic properties is the first important task for the magnetic driving
force calculation and understanding of the magneto-mechanical behavior of
Ni-Mn-Ga and similar type alloys.

Some results showing main magnetic properties of Ni$_{48}$Mn$_{30}$Ga$_{22}$
in martensitic state are indicated in Figs.3-4. First, the magnetization
behavior of a single variant of martensitic phase has been studied. For this
aim the internally twinned sample was initially transformed into a single
variant state by application of compressive stress $\left( >3MPa\right) $
and then hardly constrained without unloading in a special holder. Such a
procedure was needed to avoid the magnetic field induced motion of twin
boundaries and resulting transformation between the different twin variants
during the magnetization measurements. Two magnetization loops displaying
field dependence of the magnetization $m_a\left( h\right) $ along the easy
(tetragonal symmetry) axis and also along the hard magnetization direction $%
m_t\left( h\right) $ are shown in Fig.3. Different mechanisms of
magnetization for easy and hard magnetization directions are expected,
respectively. In particular, 180$^0$ magnetic domain walls motion is
expected to be the main magnetization mechanism for easy axis. So as, the
rotation of magnetization without domain walls motion seems to be
responsible for the magnetization behavior in a hard direction [5].
Corresponding magnetization free energies can be calculated from these
magnetization data as functions of the magnetic field applied as follows:

\begin{equation}
\label{eq1}g_a\left( h\right) =-\int\limits_0^hm_a\left( h\right)
dh;\,\,\,\,\,\,\,\,\,\,\,\,\,\,g_t\left( h\right) =-\int\limits_0^hm_t\left(
h\right) dh; 
\end{equation}

These results are shown in Fig.4. The energy of magnetic anisotropy defined
as a difference of these free energies is also indicated on this plot. In
particular, it can be easily found that the magnetic driving force
responsible for twin boundaries motion is practically equal to the magnetic
anisotropy energy. For this aim one should also know the magnetization free
energy in the multi-variant sate. Its dependence on the relative volume
fractions of different martensite variants is especially important.

Fig.5, for instance, schematically shows a typical for some new Ni-Mn-Ga
alloys two-variant twin microstructure, orientation of easy and hard
magnetization axes for both twin variants and also the magnetic driving
force alignment that usually give the maximal value of the magnetic
field-induced strain in this material. In this case of simple two-variant
twinning geometry the easy magnetization axis of the first variant (white
area) is parallel to a magnetic field applied. For the second one (gray
color) the magnetic field is applied in a transversal hard magnetization
direction. Further, both these variants are called axial $(a)$ and
transversal $(t)$, respectively. Taking the specific magnetization free
energies for both twin variants $g_a\left( h\right) $ and $g_t\left(
h\right) $ found, for instance, from the experiment one can write the total
magnetization free energy per unit volume of the material as follows:

\begin{equation}
\label{eq2}g_{mag}\left( h,x\right) =xg_a\left( h\right) +\left( 1-x\right)
\,g_t\left( h\right) 
\end{equation}
Therefore, due to magnetic anisotropy the magnetization free energy becomes
dependent on the relative volume fractions $x$ and $1-x$ occupied by axial
and transversal twin variants, correspondingly. According to our previous
calculation of the magnetization free energies shown in Fig.4 $g_a\left(
h\right) $ $<$ $g_t\left( h\right) $. So, one can decrease from its maximal
value at $x=0$ to the minimal one at $x=1$ moving the twin boundaries as
indicated in Fig.5. The corresponding magnetic driving force moving twin
boundaries along their normal directions can be found from the general
thermodynamic rule as follows:

\begin{equation}
\label{eq3}f_{mag}\left( h\right) =-\left[ \frac \partial {\partial x}%
g_{mag}\left( h,x\right) \right] _h=\,g_t\left( h\right) -g_a\left( h\right) 
\end{equation}
Therefore, a non-zero magnetic driving force responsible for twin boundary
motion appears as a result of uniaxial magnetic anisotropy of Ni-Mn-Ga. This
force is applied along the twin boundary normal direction and equal to a
difference in magnetization free energies between the different twin
variants. The magnetic driving force is dependent on the magnetic field
applied and never can exceed its saturation level as shown in Fig.4. The
maximal estimated value about of $0.13*10^6$ $N/m^2$ is achieved at $h>0.8T$
for Ni$_{48}$Mn$_{30}$Ga$_{22}$ alloy. Similar way, twin boundaries can be
also driven by the mechanical stress applied. The corresponding normal
mechanical driving force as a function of the applied stress $\widehat{%
\sigma }$ is known can be found as follows:

\begin{equation}
\label{eq4}f_{mec}\left( \widehat{\sigma }\right) =\widehat{\varepsilon }_0%
\widehat{\sigma }=\varepsilon _0\left( \sigma _{xx}-\sigma _{yy}\right) 
\end{equation}
Here, $\widehat{\varepsilon }_0$ is the strain matrix associated with
twinning transformation between two single variants of martensitic phase. In
the $XY$ reference system shown in Fig.5 it is represented as a diagonal
matrix as follows:

\begin{equation}
\label{eq5}\widehat{\varepsilon }_0=\left( 
\begin{array}{ccc}
\varepsilon _0 & 0 & 0 \\ 
0 & -\varepsilon _0 & 0 \\ 
0 & 0 & 0 
\end{array}
\right) 
\end{equation}
where, $\varepsilon _0=1-c/a=0.0579$ is the twinning transformation strain
value that can be easily estimated from the martensitic phase lattice
parameters data: $a\simeq b=0.594nm$ and $c=0.562nm$. This is a maximal
strain allowed by crystallography of twinning in Ni-Mn-Ga.

\section{Universality rules and relationship between the stress-induced and
magnetic-field-controlled twinning strains}

No general way has been found up to date that would allow calculate the
macroscopic twinning strain as a function of driving force. Nevertheless,
one can expect that independently on the physical nature of these forces
(magnetic, mechanical, etc.) the same driving forces applied to twin
boundaries must produce the same macroscopic deformation effects. This
evident physical requirement or universality rule denotes that the
macroscopic twinning strain must be some universal function dependent only
on the driving force value and independent on the physical source of this
force. So, in case of plane twinning both non-zero components of the strain
can be written through a single universal function $\varepsilon ^u\left(
f\right) $ dependent on the normal driving force $f$ as follows:

\begin{equation}
\label{eq6}\varepsilon _{yy}=-\varepsilon _{xx}=\varepsilon ^u\left(
f\right) 
\end{equation}
Using this equation one can represent both the mechanical and magnetic
strain simply assuming $f$ to equal to $f_{mec}\left( \widehat{\sigma }%
\right) $ or $f_{mag}\left( h\right) $ respectively:

\begin{equation}
\label{eq7}\varepsilon _{yy}^{mec}\left( \widehat{\sigma }\right)
=-\varepsilon _{xx}^{mec}\left( \widehat{\sigma }\right) =\varepsilon
^u\left( \widehat{\varepsilon }_0\widehat{\sigma }\right) 
\end{equation}
and

\begin{equation}
\label{eq8}\varepsilon _{yy}^{mag}\left( h\right) =-\varepsilon
_{xx}^{mag}\left( h\right) =\varepsilon ^u\left( g_t\left( h\right)
-g_a\left( h\right) \right) 
\end{equation}
It is not difficult to eliminate the unknown universal function $\varepsilon
^u\left( f\right) $ from Eqns.(7,8) and find that $\widehat{\varepsilon }%
_{}^{mag}\left( h\right) $ and $\widehat{\varepsilon }_{}^{mec}\left( 
\widehat{\sigma }\right) $will take equal values if only the corresponding
mechanical and magnetic driving forces are also equal. In other words:

\begin{equation}
\label{eq9}\widehat{\varepsilon }_{}^{mag}\left( h\right) =\widehat{%
\varepsilon }_{}^{mec}\left( \widehat{\sigma }\right)
,\,\,\,\,\,\,if\,:\,\,f_{}^{mec}\left( \widehat{\sigma }\right)
=f_{}^{mag}\left( h\right) 
\end{equation}
For instance, in case of the uniaxial compressive stress $(\sigma
_{xx}=\sigma _{zz}=0;\,\,\,\,\,\sigma _{yy}=-\sigma )$ the mechanical and
magnetic driving forces become equal at:

\begin{equation}
\label{eq10}\sigma =\sigma _{mag}\left( h\right) =\varepsilon _0^{-1}\left(
g_t\left( h\right) -g_a\left( h\right) \right) 
\end{equation}
This immediately gives the following important relationship:

\begin{equation}
\label{eq11}\widehat{\varepsilon }_{}^{mag}\left( h\right) =\widehat{%
\varepsilon }_{}^{mec}\left( \sigma _{mag}\left( h\right) \right) 
\end{equation}
which allows performing the quantitative calculations of the
magnetic-field-controlled deformation effects in different ferromagnetic
shape memory alloys by using the corresponding mechanical testing data.
According to Eq.(11) the deformation effect of a magnetic field is
equivalent to some additional uniaxial compressive stress applied. This
equivalent magnetic stress $\sigma _{mag}\left( h\right) $ can be easily
calculated from the Eq.(10). For this aim one needs only to know the
magnetization properties of the material and use the magnetic driving force
calculation data indicated in Fig.3-4. The maximal equivalent magnetic
stress that can be developed in a magnetic field is at about 2.25 MPa in Ni$%
_{48}$Mn$_{30}$Ga$_{22}$.

On the other side, according to results of the mechanical testing
experiments shown in Fig.6 the uniaxial mechanical stress required to
transform completely Ni$_{48}$Mn$_{30}$Ga$_{22}$ samples from one single
variant of martensitic phase to another one through the stress-induced twin
boundaries motion does not exceed 2.5 MPa. The maximal compression strain
that can be achieved on this load is about 5.8\%. Therefore, due to very low
twinning stress value recently found in some Ni-Mn-Ga alloys the calculated
magnetic stress is completely enough to achieve very high 5\% magneto-strain
value in these new materials.

\section{Model calculations}

Some calculation results that follows from our model consideration are
presented in this section. In particular, Fig.7 represents both the
strain-stress hysteresis loop found from the experiment and also a set of
model fitting curves corresponding to different values of twinning stress.
Simple Fermi-like distribution functions were used as an appropriate fitting
basis for analytic interpolation of the mechanical testing results.

\begin{equation}
\label{eq12}\widehat{\varepsilon }_{\pm }^{mec}\left( \sigma \right)
=\varepsilon _0\left( 1+\exp \left( \frac{\pm \sigma _0-\sigma }{\Delta
\sigma }\right) \right) ^{-1} 
\end{equation}
Here, $\pm $ denote loading and unloading curves respectively. $\sigma _0$
and $\Delta \sigma $ are characteristic stress parameters that can be
associated with the start $\left( \sigma _s=\sigma _0-2\Delta \sigma \right) 
$ and finish $\left( \sigma _f=\sigma _0+2\Delta \sigma \right) $ twinning
stress values. The best fit between the model function and experimental data
is achieved at $\sigma _0=1.56\,MPa$ and $\Delta \sigma =0.26\,MPa$ and
indicated in Fig.7 by the solid circle labelled line. Additional mechanical
hysteresis loops $(a-b,c-e)$ shown here were also simulated to study the
effect of different twinning stress values on the magneto-strain behavior.

The corresponding model calculation results performed in accordance with
Eq.(11) and the experimental measurement data for the
magnetic-field-controlled strain behavior in Ni$_{48}$Mn$_{30}$Ga$_{22}$ are
plotted altogether in Fig.8. As follows from these calculations the increase
of twinning stress may cause the gradual decrease of the magnetic field
induced strain from its maximal possible value 5.8\% at $\sigma _f<1.67\,MPa$
to 2\% at $\sigma _f=3.32\,\,MPa$. In particular, one can observe the
reasonable quantitative agreement between the 5.5\% value expected from the
model calculation and 5.2\% experimental one. Therefore, one can conclude
that the low twinning stress value is a very critical physical parameter
controlling a super-large magneto-strain effect in Ni-Mn-Ga. In particular,
this can explain considerably lower magneto-strain effect found earlier in
some nearly stoichiometric alloys which had at almost ten times higher$%
(15-20MPa)$ twinning stresses. So, the magnetic driving forces (that can
produce no more than $2-3MPa$ equivalent stress) were only able
insufficiently redistribute the relative twin variant fractions in these
materials. As a result no more than 0.2\% deformation effect was observed.

The next Fig.9 represents another important application of the model
concerning one interesting physical effect that has been found in
magnetization behavior. This effect is also directly connected with the
magnetic-field-controlled twin boundary motion found in Ni-Mn-Ga alloy. Some
very special magnetization hysteresis loop is observed if the magnetic field
is applied to a single variant of Ni-Mn-Ga alloy along its hard
magnetization direction. At the initial stage the magnetization curve
exactly coincides with a corresponding hard magnetization curve shown in
Fig.(3). At higher magnetic field the sample is transformed by twinning
until the second most appropriate twin variant with easy axis parallel to a
magnetic field applied will be formed. During the next magnetic cycles the
magnetization curve will always follow the easy magnetization path
practically without any hysteresis. The quantitative calculation of this
effect can be done according to the following magnetization equation:

\begin{equation}
\label{eq13}m\left( \varepsilon ,h\right) =m_t\left( h\right) +\left(
\varepsilon /\varepsilon _0\right) \left( m_a\left( h\right) -m_t\left(
h\right) \right) 
\end{equation}
that has been first found in [3] and can also be easily obtained in the
framework of our present approach by using the well known thermodynamic
definition:

\begin{equation}
\label{eq14}m\left( \varepsilon ,h\right) =\left( -\frac \partial {\partial h%
}g_{mag}\left( h,x\right) \right) _{x=\varepsilon /\varepsilon _0} 
\end{equation}
and evident relationship $x=\varepsilon /\varepsilon _0$ between the strain
and transformed fraction. Resulting behavior of the magnetization can be
found from the Eq.(13) and previously calculated field dependence of the
magneto-strain, assuming there $\varepsilon =$ $\varepsilon _{}^{mag}\left(
h\right) $. Calculations were also done for different twinning stress
values. This effect shows also a good agreement between the model and
experiment. Finally, one should make the following important conclusions:

\section{Conclusions}

1. A super-large deformation effect at about 5\% that can be induced by
application of magnetic field less than $0.8T$ in some new ferromagnetic
non-stoichiometric Ni-Mn-Ga shape memory alloys seems to be the most
interesting property of this material.

2. Ni-Mn-Ga alloys displaying giant field-induced deformation effects have
always very simple practically two-variant twin microstructure. As a result,
these alloys can be easily deformed by application of the low $<2MPa$
mechanical stress or magnetic field less $0.8T$ through the mechanism of
twin boundary motion. In particular, the reversible magneto-mechanical
transformation cycle between two twin variants of martensitic phase can be
induced by subsequent application of magnetic field along two different
perpendicular directions.

3. The magnetic driving force applied to twin boundaries is equal to a
difference in magnetization free energies between the different twin
variants of martensite. This difference also characterizes the energy of
uniaxial magnetization anisotropy and can be calculated from the
corresponding magnetization measurements. The magnetic driving force
achieves its maximal value 0.13 MN/m2 in a magnetic field higher $0.8T$. In
agreement with our earlier estimations and present model calculations this
value is completely enough to explain high 5\% magnetic field induced strain
in some Ni-Mn-Ga alloys which have the very low $<2MPa$ twinning stress.

4. There is a definite analogy between the deformation effects caused by the
mechanical and magnetic driving forces. For instance, in both cases the
macroscopic twining strain driven by the mechanical stress or the magnetic
field applied can be expressed through the same universal function dependent
on the corresponding (mechanical or magnetic) driving force. This
universality rule allows to perform the quantitative calculations of the
magnetic field induced strain as a function of field by using the mechanical
testing results.

\newpage Figure cuptions

Fig.1. Giant compression and extension magneto-strain effects induced by
magnetic field in Ni$_{48}$Mn$_{30}$Ga$_{22}$ alloy.

Fig.2. Different stages of twin boundary motion in situ observed in the
rotating magnetic field 0.6T in Ni$_{48}$Mn$_{30}$Ga$_{22}$ alloy. 

Fig.3. Magnetization curves along easy and hard axes of Ni$_{48}$Mn$_{30}$Ga$%
_{22}$ sample constrained in a single variant state. 

Fig.4. Magnetic anisotropy of Ni$_{48}$Mn$_{30}$Ga$_{22}$ and field
dependence of magnetization free energies for easy and hard magnetization
directions 

Fig.5. Two-variant twin microstructure, easy magnetization axes alignment
and magnetic driving force responsible for twin boundary motion in a
magnetic field applied. 

Fig.6. Data of mechanical testing for Ni$_{48}$Mn$_{30}$Ga$_{22}$ sample
during its transformation between two twin variants of martensitic phase. 

Fig.7. Strain-stress behavior caused by transformation between two single
twin variants induced by mechanical stress in Ni-Mn-Ga martensite. 

Fig.8. Model calculation result of the magnetic field induced strain effect
in comparison with experiment. 

Fig.9 Experimental data and model calculation of magnetization hysteresis
caused by magnetic field controlled motion of twin boundaries in Ni-Mn-Ga.

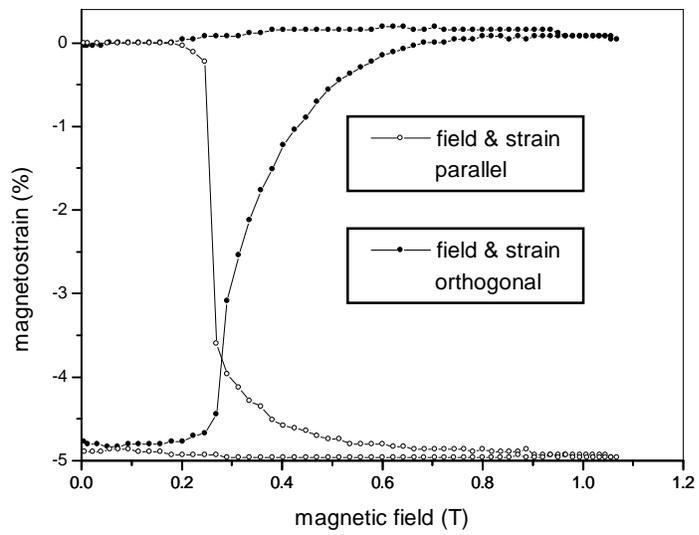
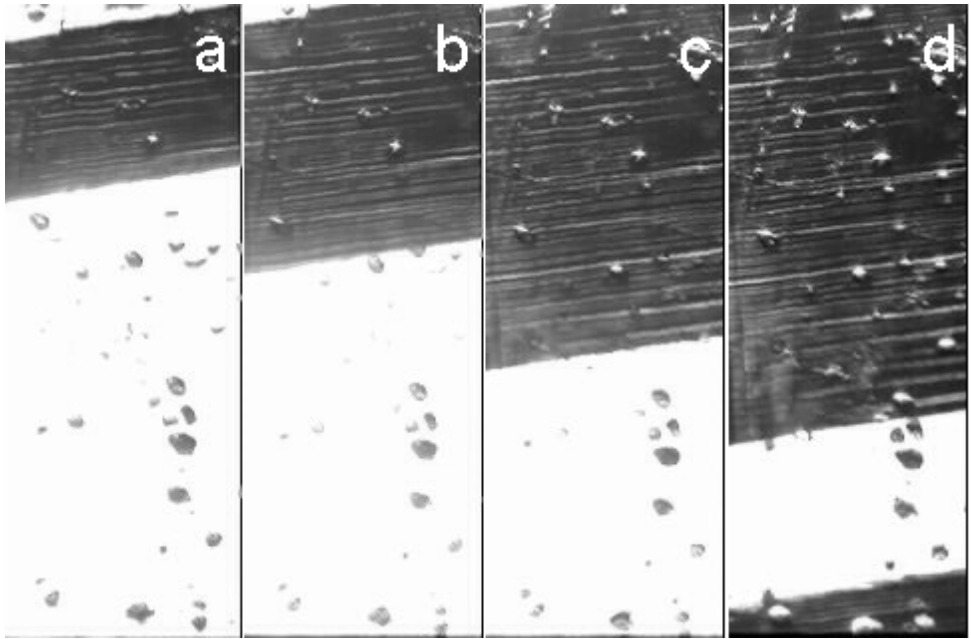
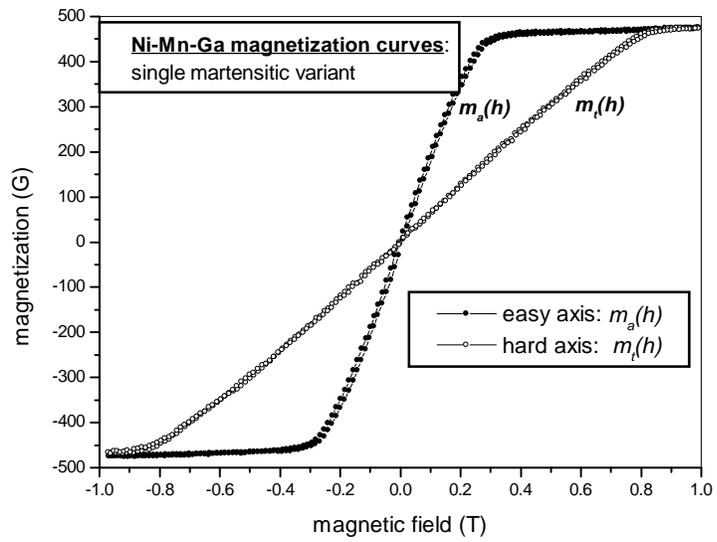
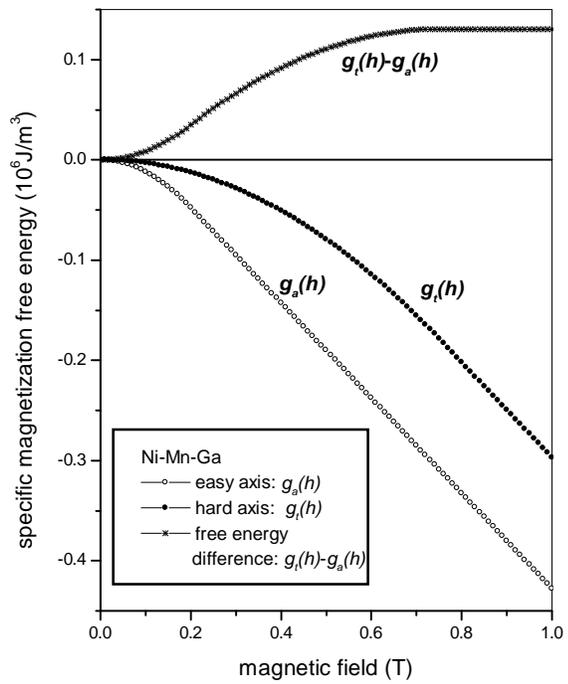
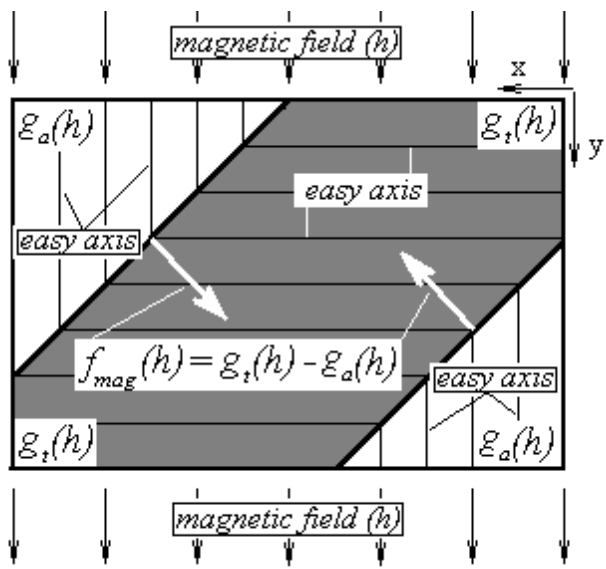
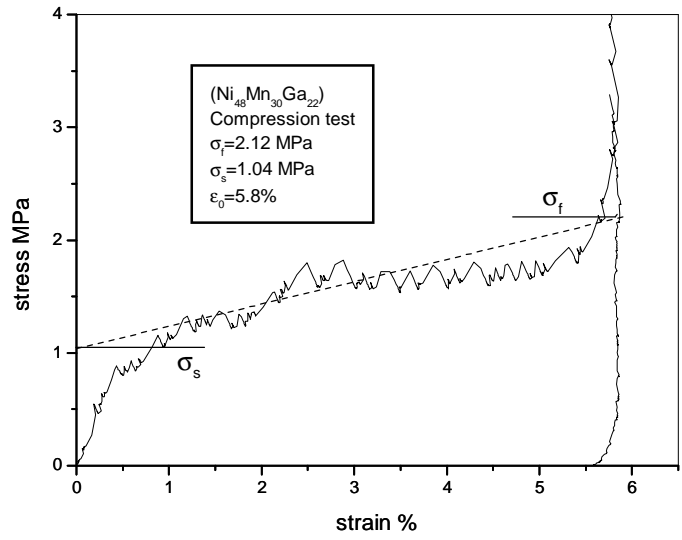
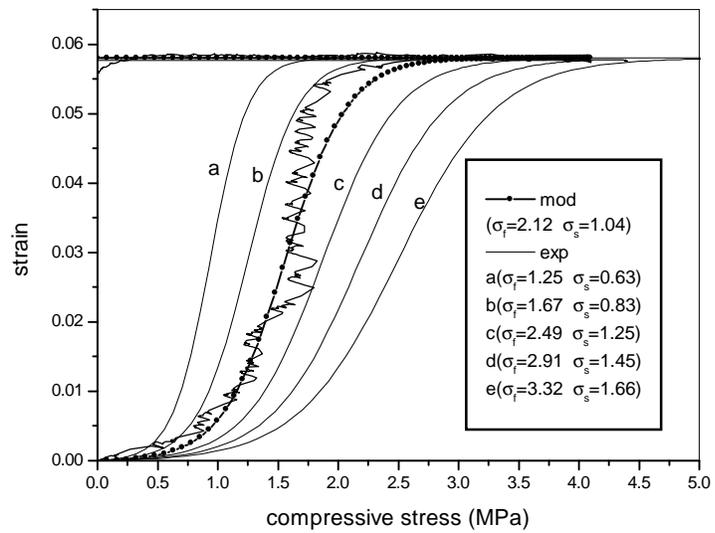
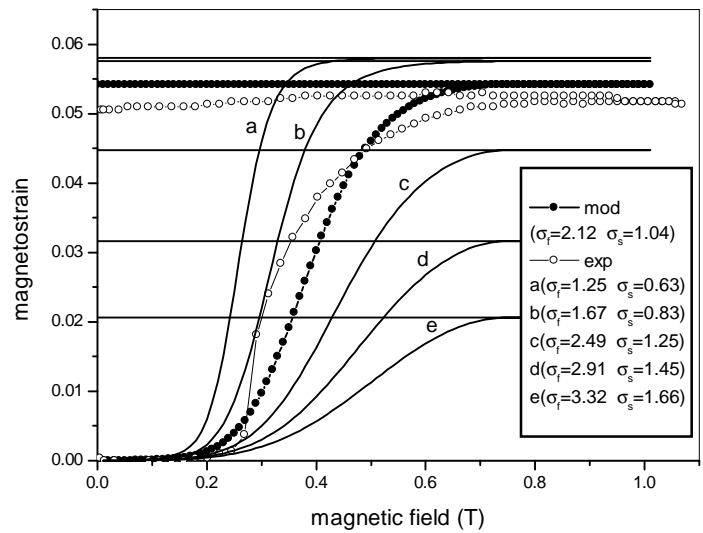
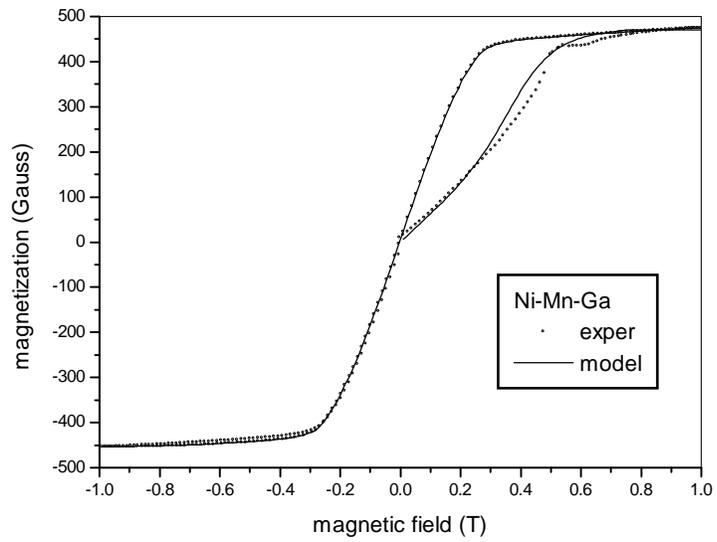
\end{document}